\documentclass[%
groupedaddress,
 amsmath,amssymb,
 aps,
prl,
twocolumn,
floatfix,
notitlepage
]{revtex4-1}

\usepackage{graphicx}% Include figure files
\usepackage{times}
\usepackage{dcolumn}% Align table columns on decimal point
\usepackage{bm}% bold math
\usepackage{subfigure}%subfigure enviroment
\makeatletter
\renewcommand{\thesubfigure}{(\alph{subfigure})}% (a) -> a or a 0 -> (a)
\renewcommand{\@thesubfigure}{\thesubfigure\hskip\subfiglabelskip}% a -> a)
\makeatother
\usepackage[colorlinks=true,bookmarks=true,%
     pdfborder={0 0 0},linkcolor=blue,urlcolor=blue,citecolor=blue,%
     breaklinks=true,bookmarksnumbered=true]{hyperref}
\def \x{\bm{x}}

\def \S{\bm{S}}

\def \k{\bm{k}}

\def \pt{\partial}

\def \L{\mathcal{L}}
\newcommand{\abs}[1]{\lvert#1\rvert}

\newcommand{\ev}[1]{\mbox{$\langle #1 \rangle$}}

\newcommand{\ua}{\uparrow}
\newcommand{\da}{\downarrow}

\begin{document}

\title{Effective Exchange Interactions for Bad Metals  and Implications for Iron-based Superconductors
}% Force line breaks with \\

\author{Wenxin Ding$^{1,2}$}\email{wenxinding@gmail.com}
 \author{Rong Yu$^3$}
 \author{Qimiao Si$^2$}
 \author{Elihu Abrahams$^4$}
 \email{deceased}
 \affiliation{%
   $^1$School of Physics and Material Science, Anhui University, Anhui Province, Hefei, 230601, China\\
$^2$Department of Physics \& Astronomy, Rice University, Houston, Texas 77005, USA\\
$^3$Physics Department and Beijing Key Laboratory of
Opto-electronic Functional Materials and Micro-nano Devices, Renmin University, Beijing 100872, China\\
 $^4$Department of Physics and Astronomy, University of California Los Angeles, Los Angeles, California 90095, USA }%
%\author{Elihu Abrahams}

\date{\today}% It is always \today, today,

\begin{abstract}
The experimentally observed bad metal behavior in parent iron pnictides and chalcogenides suggests
that these systems contain strong electronic correlations and are on the verge of a metal-to-insulator transition.
The magnetic excitations in this bad-metal regime mainly derive from the incoherent part of the electronic spectrum
away from the Fermi energy.
We present a microscopic study of the exchange interactions in such a regime within a slave rotor approach.
%We show how the effective exchange interactions can be determined not only for the localized side
%(where the coherent spectral weight $Z$ vanishes) but also in the bad-metal regime (where $Z$ is small but nonzero).
We find that the exchange interaction is maximized near the Mott transition.
Generalizations to the multi-orbital case  are discussed, as are the implications for the strength
of superconducting pairing amplitude in iron-based superconductors.
%\begin{description}
%\end{description}
\end{abstract}

\pacs{Valid PACS appear here}% PACS, the Physics and Astronomy
\maketitle
%\tableofcontents

%\input sec1.tex
{\it Introduction:~~}
Superconductivity in the iron pnictides and chalcogenides occurs at the border of antiferromagnetic order
\cite{Kamihara2008,DelaCruz2008}.
For an understanding of the superconductivity, it is important to characterize the magnetism.
An important clue for the latter is that the parent iron pnictides are bad metals.
Their electrical resistivity at room temperature is very large,
reaching the Mott-Ioffe-Regel limit ($k_F \ell < 2 \pi$) \cite{Abrahams2011a,Hussey2004}.
Optical conductivity measurements show a large suppression of the Drude weight \cite{Qazilbash2009}, which suggests that
the majority of the electronic excitations lives in the incoherent part away from the Fermi energy and the system
is in proximity to a Mott insulator \cite{Si2008,Si2009a,Haule2008}. The role of the correlation effects is further
highlighted by the observation of both the insulating states \cite{Fang2011,Gao2014,Yu2011b,Dagotto2013,Zhu2010}
and an orbital-selective Mott phase \cite{Yi2013,Yu2013a}
in a number of iron chalcogenides
 and it has also been emphasized from
a variety of perspectives \cite{Fang2008,Xu2008,Ma2008,Han2009,Bascones2012b,Seo2008,Chen2009,Moreo2009,Lv2010,Uhrig2009,Ishida2010,Giovannetti2011,Bascones2012}.

When the majority of the single-particle excitations are incoherent, they give rise to quasi-localized moments,
which are coupled with each other through frustrating exchange interactions \cite{Si2008,Yildirim2008,Ma2008,Han2009,Bascones2012b}.
This provides a natural basis to understand the large spin spectral weight observed
in both the iron pnictides\cite{Wang2011a} and iron chalcogenides \cite{Bao2011, Zhao2012,Free2010, McCabe2014}.

In this paper, we study the exchange interactions in the bad-metal regime.
While it is standard to derive superexchange interactions
in the Mott localized regime, the microscopic basis for the exchange interactions in the regime of bad metals
is much less understood.
Here we show how such exchange interactions can be derived in a microscopic framework,
using the slave-rotor approach\cite{Florens2002,Florens2004}. Important for our analysis is that this approach already contains
incoherent excitation spectra at the saddle-point level. We show
how such incoherent spectra can be integrated out to yield an exchange interaction,
not only for the localized side but also in the bad-metal regime. As a consequence, we show that
 the exchange interaction is maximized near the Mott transition.

{\it Slave-Rotor Approach:~~}
We
consider
the Hubbard model on a square lattice with only nearest-neighbor hopping
\begin{equation}
  \label{eq:Hubbard-model}
  H_{HM}( d ) = \sum_{i} H_{at}(i) - \sum_{ij,\alpha} (t_{ij} d^{\dagger}_{i\alpha} d_{j\alpha} + h.c.),
\end{equation}
in which $H_{at}(i) =
%\sum_{\alpha} \epsilon_d d^{\dagger}_{i,\alpha} d_{i,\alpha} +
\frac{U}{2}\Big(\sum_\alpha d^{\dagger}_{i\alpha} d_{i\alpha} - N/2 \Big)^2,$ $\alpha$ is
the spin/orbital index running from  $\alpha = 1,\dots, N$,
with N=2 for the one-band model.
For definiteness, we will consider a square lattice,
and only hopping between nearest-neighbor sites, $<ij>$.
It is realized that the full energy spectrum of $H_{at}(i)$ can be economically represented by a rotor kinetic energy $H_{at}(i) \rightarrow U \hat{L}_{i}^2 /2$ \cite{Florens2002,Florens2004} with $\hat{L}_{i} = -i\pt_{\theta_i}$, thus providing a tractable reference point for perturbative expansion in $t/U$. Then in this slave-rotor representation, the bare electron operator is written as a product of the auxiliary rotor fields and a fermionic spinon operator
 $d_{i\alpha} \equiv f_{i\alpha} e^{-i \theta_i}$,
with the constraint
%\begin{equation}
%  \label{eq:contraint1}
$ \hat{L_i} = \sum_{\alpha}(f^{\dagger}_{i\alpha} f_{i\alpha} -1/2) .$
%\end{equation}
In place of the phase field  one could work with the complex field
$e^{i\theta_i} = X_i$, with the additional constraint
%\begin{equation}
 % \label{eq:contraint2}
$\abs{X_i}^2 = 1 .$
%\end{equation}
The two constraints are enforced by introducing two Lagrangian multipliers, $h_i$ and $\lambda_i$.
In terms of the fermionic $f_i$ and  complex rotor $X_i$ operators, the physical $d_i$-electron operator at site $i$ is
expressed
as follows:
\begin{equation}
  \label{eq:d_by_fX}
  d_{i\alpha} \equiv f_{i\alpha} X_i^{*} \, .
\end{equation}
  
A saddle point arises when one generalizes each $X_i$ to M species so that its symmetry becomes $O(2M)$, scales the hopping
$t_{ij}$ to $1/M$, and take the large (N,M) limit with a fixed ratio $M/N$. In our analysis below,
we will write our equations for $N=2M=2$.

Using $\pt_\tau \theta_i = -i X_i^* \pt_\tau X_i$, we have the Lagrangian
\begin{equation}\label{eq:mf-L}
  \begin{split}
    & \L_{HM}  = \sum_{i,\alpha} f^\dagger_{i\alpha} (\pt_\tau  + h_i) f_{i\alpha} + \sum_i \Big(\frac{\abs{\pt_\tau X_i}^2}{U}  \\
    & + \frac{h_i}{U}(X_i^* \pt_\tau X_i - h.c.) + \lambda_i (\abs{X_i}^2 - 1)\Big)
\\ & +  \sum_{ij,\alpha} ( t_{ij} f^\dagger_{i,\alpha} f_{j,\alpha} X_i X^*_j + h.c.).
  \end{split}
\end{equation}
Note that $\frac{U}{2} \sum_{i} \hat{L}_{i}^2 =  \frac{\abs{\pt_\tau X_i}^2}{2U}$; we have rescaled
 $U$ to $U/2$ in Eq. (\ref{eq:mf-L}) to preserve the correct atomic limit \cite{Florens2002}.

The saddle point  \cite{Florens2004,Lee2005} corresponds to
decoupling the spinon-boson coupling term via
\begin{equation}
  Q_{f,ij}  = \ev{ X_j^* X_i}, \qquad
  Q_{X,ij}  = \ev{\sum_\alpha f_{j\alpha}^\dagger f_{i\alpha} },
\end{equation}
which can be formally done by a Hubbard-Stratonovich transformation. These are the channels that preserve physical fermion numbers.
The Lagrangian $\L_{HM}$ is decoupled into two parts:
\begin{equation}
  \label{eq:mf-Lf}
  \L_{MF,f} = \sum_{i,\alpha} f^\dagger_{i\alpha} (\pt_\tau + h_i) f_{i\alpha} + t \sum_{\ev{ij},\alpha}
  (Q_{f} f^\dagger_{i\alpha} f_{j\alpha} + h.c.), 
\end{equation}
\begin{equation}
  \label{eq:mf-LX}
  \begin{split}
    \L_{MF,X} & = \sum_i \Big(\frac{\abs{\pt_\tau X_i}^2}{U} +
    \frac{h}{U}(X_i^* \pt_\tau X_i - h.c.) \\ &+ \lambda_i\abs{X_i}^2
    \Big) + t \sum_{\ev{ij}} (Q_{X} X_i X^*_j + h.c.).
  \end{split}
\end{equation}
The diagrams shown in Fig. \ref{subfig:mf-diagram} correspond to the saddle point (see below).
From here on, we drop the $_{ij}$ index for $t$, keeping only the nearest neighbor hopping, and also that for $Q_{f(X)}$, assuming translational invariance throughout this paper.
Then the spinon and $X$-field Green's functions at the saddle-point level read
\begin{equation} \label{eq:Gf-MF}
  G_f(\omega;\k) = (i \omega + h
  -Q_f \epsilon_{\k})^{-1},
\end{equation}
\begin{equation}
  \label{eq:GX-MF}
  G_{X}(\nu;\k) = (\nu^2/U + 2ih\nu/U + \lambda + Q_X \epsilon_{\k})^{-1},
\end{equation}
where
$\epsilon_{\k} = -2 t [\cos(k_x) + \cos(k_y)]$
is the bare lattice dispersion function.
For the saddle point solution, the Lagrangian multipliers become uniform: $h_i \rightarrow h$ and $\lambda_i \rightarrow \lambda$.
The self-consistent equation which determines $\lambda$ reads
\begin{equation}\label{eq:sc-lambda}
\int \frac{d^2k}{(2\pi)^2} \sum_{\nu}G_{X} (\nu;\k) = 1.
\end{equation}
The $h$ is determined by $\ev{\hat{L}} = \sum_\alpha(\ev{f^\dagger_{i\alpha} f_{i\alpha}} - 1/2)$.
For the half-filling case we consider,
$h = 0$ for arbitrary $U$.

In both the insulating phase and the metallic phase, the spinons are always treated as free fermions at half-filling.
We find $Q_X = \ev{\sum_{\alpha} f^\dagger_{i\alpha} f_{j\alpha}} = 8/\pi^2$, irrespective of $U$. Thus self-consistency is automatically satisfied. $Q_f$ in general decays with increasing $U$, and in the large-$U$ limit $Q_f \simeq 2/U$.
The Mott transition is realized when $U$ reaches $U_c$ where $(\lambda + Q_X \epsilon_{\k=0})$ vanishes,
so the $X$-field starts to condense. For $U<U_c$, we can divide the rotor field into
a condensate and an incoherent component: $X_i \rightarrow X_i^{0} + X_i'$ and,
correspondingly, the $X$-field Green's function can be written as
\begin{equation}
  \label{eq:GX-MF-metallic}
  G_X(\nu;\k) = Z \delta(\nu)\delta(\k) + G_{X,\text{inc}},
\end{equation}
where $G_{X,\text{inc}} = \ev{X^{*\prime}_{\k} X'_{\k}} = (\nu^2/U + \lambda_C + Q_X \epsilon_{\k})^{-1}$ and $Z = (X_i^{0})^2$.
In the metallic phase, $\lambda = \lambda_C$ remains a constant, determined by $\lambda_C = - Q_X \epsilon_{\k=0}$.
Then from Eq. (\ref{eq:sc-lambda}), we find
%$Z= 1 - \sqrt{U/U_c}$,
\begin{equation}\label{eq:Z}
Z= 1 - \sqrt{U/U_c} ,
\end{equation}
with $U_c$ is determined from
$\int \frac{d\nu d^2k}{(2 \pi)^3} (\nu^2/U_c +\lambda_C +Q_X \epsilon_{\k})^{-1} = 1$.
The spinon Green's function Eq. (\ref{eq:Gf-MF}) remains the same (up to the renormalization factor $Q_f$).
The division of the $d$-electron excitations into coherent and incoherent parts is thus realized by separating the rotor field $X$
into a condensate and a fluctuating part.
\begin{figure}
  \centering
  \subfigure[]{\includegraphics[width=3.3cm]{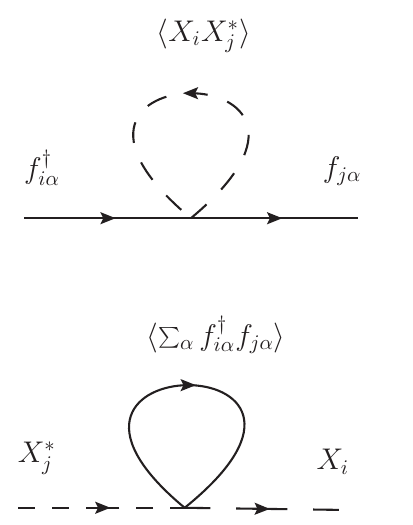}\label{subfig:mf-diagram}}
  \subfigure[]{\includegraphics[width=5.cm]{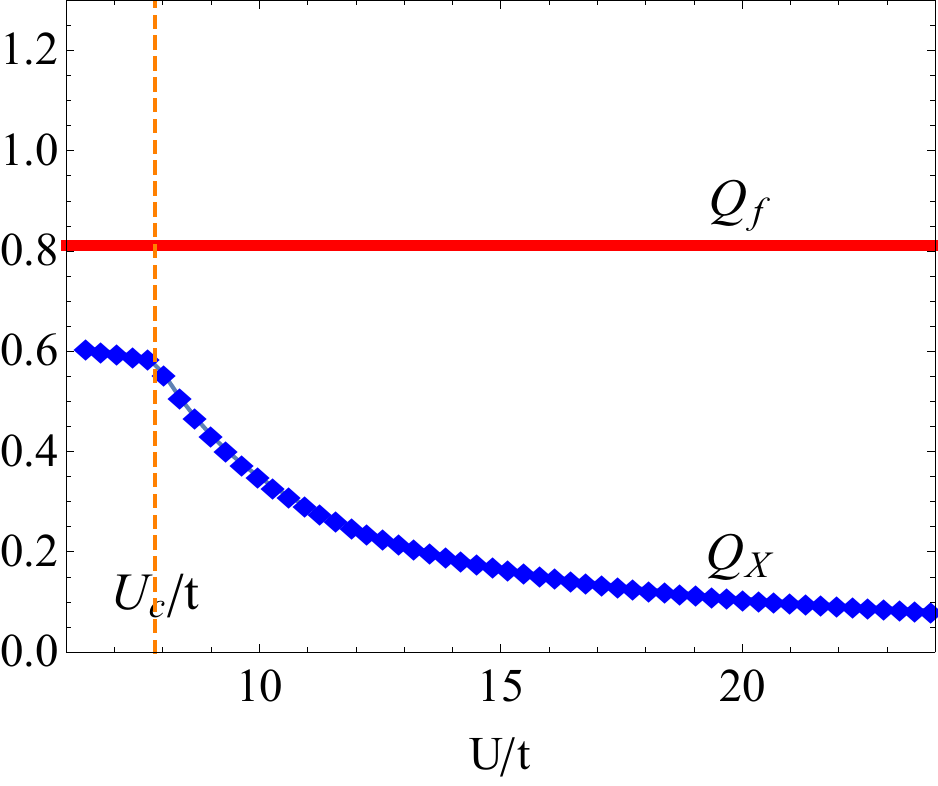}\label{subfig:mf-parameters}}
  \caption{(a) Feynman diagrams corresponding to the saddle point equations. (b) The self-consistent parameters
  $Q_f$ and $Q_X$ plotted as function of $U/t$.}\label{fig:mf}
\end{figure}
The parameters $Q_f$ and $Q_X$ computed numerically as a function of $U/t$ are
shown in Fig. (\ref{subfig:mf-parameters}).

\begin{figure}
  \centering
  \subfigure[]{\includegraphics[width=8cm]{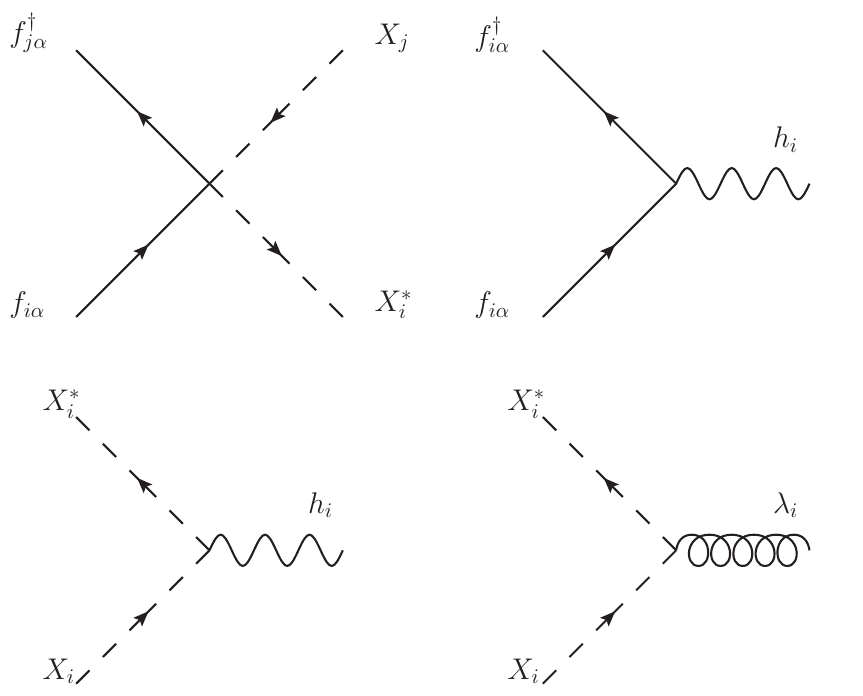}\label{subfig:bare-vortex}}
  \subfigure[]{\includegraphics[width=8cm]{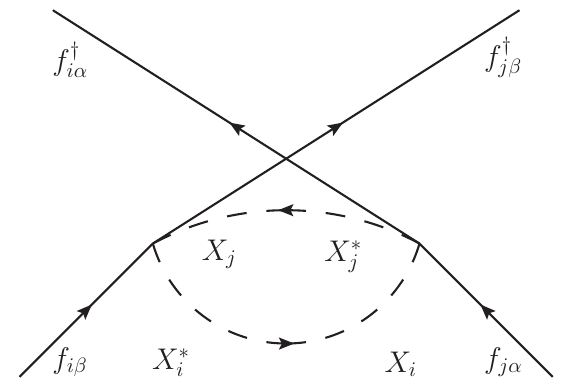}\label{subfig:Heisenberg}}
  %\subfigure[]{\includegraphics[width=8.5cm]{./Fig-3b.pdf}\label{subfig:rotor-spectral-function}}
 % \includegraphics[width=8.5cm]{./slave_rotor_spectral_func.pdf}
  \caption{(Color online) (a) Bare interaction vortices.
%The first diagram describes a  spinon-rotor vortex, which corresponds to the hopping of the physical electrons.
%The other three come from the constraints being enforced by the Lagrangian multipliers;
    (b) Feynman diagram for the effective spin exchange interaction.
}\label{fig:rotor-spectral-function}
\end{figure}

{\it Exchange Interaction from Integrating Out Incoherent Excitations:~~}
Beyond the saddle point, the spinon and rotor fields are coupled.
To introduce these couplings, we consider Eq. (\ref{eq:mf-L}) diagrammatically. $\L_{HM}$ contains various bare interaction
vortices as shown in Fig. (\ref{subfig:bare-vortex}). The most important one is the first, a spinon-rotor vortex;
it corresponds to the hopping of the physical electrons. The others come from
 the constraints being enforced by the Lagrangian multiplier fields.

%To go beyond the bare perturbative expansion, we
We first (formally) integrate out
either the spinons ($f_{i\alpha}$s) or the rotors ($X_{i}$s) from the full $\L_{HM}$
%from original slave rotor representation of the Hubbard model,
of Eq. (\ref{eq:mf-L})
to obtain effective actions $S^{\text{eff},X}(X_{i})$ or $S^{\text{eff},f}(f_{i\alpha})$:
\begin{equation} %\label{eq:S-f-eff}
  S^{\text{eff},f} =  S^{0,f} - \ln \det\Big[
  \begin{pmatrix}
   (G^{0}_{X})^{-1} & t  \sum f^\dagger_{i\alpha} f_{j\alpha}  \\ t  \sum f^\dagger_{j\alpha} f_{i\alpha} & (G^{0}_{X})^{-1}
  \end{pmatrix}_{i,j}
\Big],
\nonumber
\end{equation}
\begin{equation}\label{eq:S-X-eff}
  S^{\text{eff},X} =  S^{0,X} - \ln \det\Big[
  \begin{pmatrix}
   (G^{0}_{f})^{-1} & t  X_i^* X_j \\ t  X_j^* X_i &   (G^{0}_{f})^{-1}
  \end{pmatrix}_{i,j}
\Big],
\end{equation}
where
 \begin{align}
\label{eq:atomic}
  & S^{0,f} = \int d\tau \L^{0,f} = \int d\tau \sum_{i,\alpha}f^\dagger_{i\alpha} \pt \tau  f_{i\alpha}, \nonumber \\
  & S^{0,X}  =\int d\tau \L^{0,X} = \int d\tau \sum_{i} \Big(\frac{\abs{\pt_\tau X_i}^2}{%2
  U} + \lambda \abs{X_i}^2\Big),
\nonumber
 \end{align}
 The $\ln \det[\dots]$ can be expanded in orders of $t$
 to generate effective
 hoppings and interactions.
 We take a renormalized expansion, so $G^0_f$ and $G^0_X$ are replaced by $G_f$ and $G_X$.
To the lowest order in $t$, we only have the Feynman diagrams shown in Fig. (\ref{subfig:mf-diagram}),
which give rise to the saddle point where $G_f$ and $G_X$ are
computed self-consistently. Again, for the metallic phase, we need to decompose the rotor fields into a condensate component and a fluctuating component:
$X_i \rightarrow X'_i + X^{0}_i$.

%\dwx{
%Now we expand Eq. (\ref{eq:S-f-eff}) up to $\mathcal{O}(t^2)$ which yields
%  \begin{align}
 %   S^{eff,f} = \int d\tau \L_{eff,f} =  \int d\tau (\L_{MF,f} - H_{f;\text{ex}}),
%  \end{align}
%where $H_{f;\text{ex}}$ is generated by the exchange vortex shown in Fig. (\ref{subfig:Heisenberg}). We have
For $S^{\text{eff},f}$ at $\mathcal{O}(t^2)$, we have the exchange vortex shown in Fig. (\ref{subfig:Heisenberg}).
This diagram yields 
%a
% Heinsenberg 
a magnetic
exchange interaction,
\begin{equation} \label{eq:Jex1}
\begin{split}
     H_{f;\text{ex}} &=  \frac{J_{\text{ex}}}{2} \sum_{<ij>} \sum_{\alpha,\beta} f^\dagger_{i\alpha}f_{i\beta} f^\dagger_{j\beta} f_{j\alpha} \\
& = J_{\text{ex}} \sum_{<ij>} {\bf S}_i \cdot {\bf S}_j,
  \end{split}
\end{equation}
where
%
%\begin{align}
${\bf S}_i = (1/2 )\sum_{\alpha,\beta}f^\dagger_{i\alpha} \bm{\sigma}_{\alpha \beta} f_{i\beta} $
%\end{align}\label{eq:S-f}
%
and the factor $1/2$ is because it is the second order term of the cumulant expansion.
(We have ignored an additive constant in the last equality.) 
%\qs{Because of the constraint in the slave-rotor representation,
%${\bf S}_i$ (and likewise ${\bf S}_j$) is the same as the physical $d$-electron spin operator:
%\begin{align}
%{\bf S}_i = (1/2) \sum_{\alpha,\beta}d^\dagger_{i\alpha} \bm{\sigma}_{\alpha \beta} d_{i\beta} \, .
%\label{eq:S-d}
%\end{align}
%}

The exchange interaction is determined by the rotor bubble,
\begin{equation} \label{eq:Jex2}
\begin{split}
 J_{\text{ex}} =  \int \frac{d\nu}{2\pi}  G_{X}(\nu;i,i) G_{X}(\nu;j,j),
  \end{split}
\end{equation}
where $G_{X}(\nu;i,i) = \int \frac{d^2k}{(2\pi)^2} G_X(\nu;\k)$.
Note that, this exchange interaction operates in the spin sector, whose energies are low compared to the energies
of the incoherent poles of the slave rotors (see below). This makes the equal-time exchange interaction
to be essentially the same
as the static exchange interaction, for which Eq.~(\ref{eq:Jex1}) describes.

For the latter reference, we contrast the above with a bare perturbative expansion.
The latter is based on the following bare atomic actions and bare Green's functions of the spinons and rotors:
\begin{eqnarray}
  & G^{0}_{X}(\nu; i,j) = \delta_{ij} (\nu^2/U + \lambda)^{-1}, \nonumber\\
  &G^{0}_{f}(\omega;i,j) =  \delta_{ij}  (i \omega )^{-1} .
\end{eqnarray}
% which are the atomic actions and bare Green's functions of the spinons and rotors, respectively.
In this procedure, we can also determine an exchange interaction,
$J_{\text{ex}}^{\text{bare}}$, from Eq.~(\ref{eq:Jex2}) with $G_{X}$ replaced by $G_{X}^0$.

  {\it Relation to Physical Observables:~~}
  In the slave rotor representation, the electronic Green's function is calculated via the rotor and spinon Green's functions according to
\begin{align}
     %& i G_{d }(t;\x,\x')  =  - G_{f }(t;\x,\x')  G_X(-t;\x,\x') ~\rightarrow \\
      & i G_{d}(\omega; \k) = \nonumber \\
      &- \int d\omega' d\k' G_{f }(\omega - \omega'; \k - \k') G_X(\omega';\k'). \label{eq:Gd}
\end{align}
This is the Fourier transform of $ i G_{d }(t;\x,\x')  =  - G_{f }(t;\x,\x')  G_X(-t;\x,\x')$ which stems directly from Eq. (\ref{eq:d_by_fX}). In the Mott insulating phase, $G_d(\omega;\k)$ is fully gapped due to the large gap from $G_X(\omega;\k)$. In the bad metal regime, a coherent quasiparticle part emerges, $G_{d,coh}(\omega;\k) = Z G_f(\omega;\k)$, while the incoherent part still follows from 
Eq. (\ref{eq:Gd}) with $G_X(\omega';\k')$ replaced by $G_{X,inc}(\omega';\k')$.

For the spin operators, because of the constraint in the slave-rotor representation,
${\bf S}_i$ (and likewise ${\bf S}_j$) is the same as the physical $d$-electron spin operator:
\begin{align}
{\bf S}_i = (1/2) \sum_{\alpha,\beta}d^\dagger_{i\alpha} \bm{\sigma}_{\alpha \beta} d_{i\beta} \, .
\label{eq:S-d}
\end{align}
%Wenxin: THIS STATEMENT IS CONFUSING AND I WOULD LIKE TO SUGGEST THAT WE REMOVE IT
%This can be consistently verified by expressing the spin expectation values in terms of the physical Green's function $G_d$s.

\begin{figure}
 % \subfigure[]{\includegraphics[width=8.5cm]{./compare.pdf}\label{subfig:compare1}}
 % \subfigure[]{\includegraphics[width=8cm]{./compare2.pdf}\label{subfig:compare2}}
  \includegraphics[width=8.5cm]{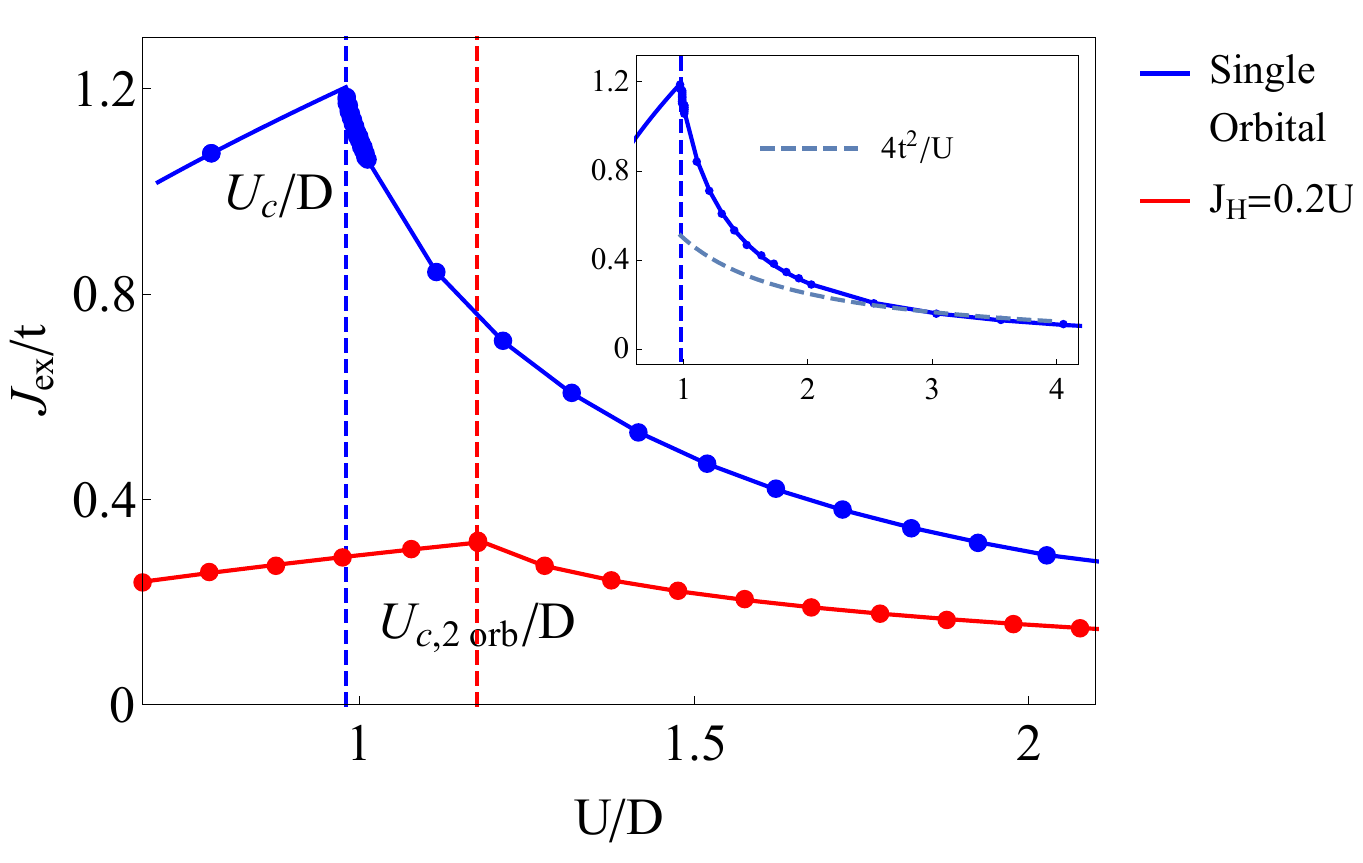}
  \caption{(Color online) The calculated $J_{\text{ex}}$ plotted as a function of $U/t$ from the bad-metal regime to the insulating side for both the single orbital case (blue) and the simple two-orbital model (red). For both cases, the exchange interactions are seen to be maximized around the Mott transition. The inset shows $J_{\text{ex}}$ for the single orbital case on the insulating side up to large values of $U/t$; the axes are the same as in the main plot. The navy dashed line corresponds to the standard result for the
  large-$U$ limit, $J_{\text{ex}} \simeq \gamma t^2/U$ with $\gamma=4$.
}\label{fig:JeffU}
\end{figure}

{\it Exchange Interaction on the Insulating Side:~~}
When $U$ is significantly larger than $U_c$,
the rotor spectrum has a large gap around $\omega = 0$,
and two peaks around $ \pm U \mathcal{O}(1)$ respectively. The latter characterizes incoherent electronic excitations,
which are responsible for $J_{\text{ex}} \sim 1/U$ behavior.
Numerical results for large $U$'s are shown in the inset of Fig. (\ref{fig:JeffU}), where we also plot the ratio
$J_{\text{ex}} / (\gamma t^2/U)$ as a function of $U$  in which $\gamma$
is determined by $J_{\text{ex}} / (\gamma t^2/U)\vert_{U\rightarrow\infty} = 1$ to compare with the standard
super-exchange interaction. Here we do find $\gamma = 4$, in agreement with the standard result.
Note that, the
behavior
at large $U$ can  be qualitatively seen by computing the rotor bubble function
with the bare Green's function of the rotors:
$ J_{\text{ex}}^{\text{bare}}(U \rightarrow \infty) =  \int \frac{d\nu}{2\pi}G^0_{X}(i \nu;i,i) G^0_{X}(i \nu;j,j)$.
Using $\lambda_{U\rightarrow\infty} = U/4$
determined from Eq. (\ref{eq:sc-lambda}), we find
$ J_{\text{ex}}^{\text{bare}} = \frac{\gamma t^2}{U}$.
Though $\gamma = 2$ (as opposed to $4$ above),
$ J_{\text{ex}}^{\text{bare}} $ does capture the $t^2/U$ dependence.

{\it Exchange Interaction in the Bad Metal Regime and across the Metal-Insulator Transition:~~}
When $U$ approaches $U_c$ from above, the evolution of the rotor spectral function (integrated over $\k$)
is illustrated by the results shown in Fig. (\ref{fig:rotor-spectral-function}) for $U/U_c=1.25, 1.1$.
The incoherent peaks are still well-defined, but the peak locations naturally
shift towards smaller $\omega$. Therefore $J_{\text{ex}}$ increases as $U \rightarrow U_c^+$.

Moving into the bad-metal regime, where $U < \sim U_c$, the coherent electron weight $Z$ is non-zero but still small.
Importantly, the incoherent peaks remain in the rotor spectral function, as illustrated by the results shown
in Fig. (\ref{fig:rotor-spectral-function})
for $U / U_c =  0.9, 0.75, 0.5$. We still have well-defined exchange interaction, $J_{\text{ex}}$
from integrating out the incoherent spectra. Importantly,
\begin{equation}
  \begin{split}
    J_{\text{ex}} & = \int \frac{d\nu}{2\pi} G_{X,\text{inc}}(\nu;i,i) G_{X,\text{inc}}(\nu;j,j)  \\
    & = \sqrt{\frac{U}{U_c}} \int \frac{d\nu'}{2\pi} G_{X,c}(\nu';i,i) G_{X,c}(\nu';j,j)\\
    & =  J_{\text{ex}}(U_c) (1 - Z),
  \end{split}
\end{equation}
where $G_{X,c}(\nu';\k) = (\nu^{\prime 2}/U_c + \lambda_c + Q_X \epsilon_{\k})^{-1}$ is transformed
from $G_{X,\text{inc}}(\nu;\k)$ by letting $\nu' = \nu \sqrt{U_c/U}$.
Because the spectral weight in the incoherent part is lost to the coherent part,
the exchange interaction will decrease as $U$ decreases from $U_c$.
\begin{figure}
  \centering
\includegraphics[width=8.5cm]{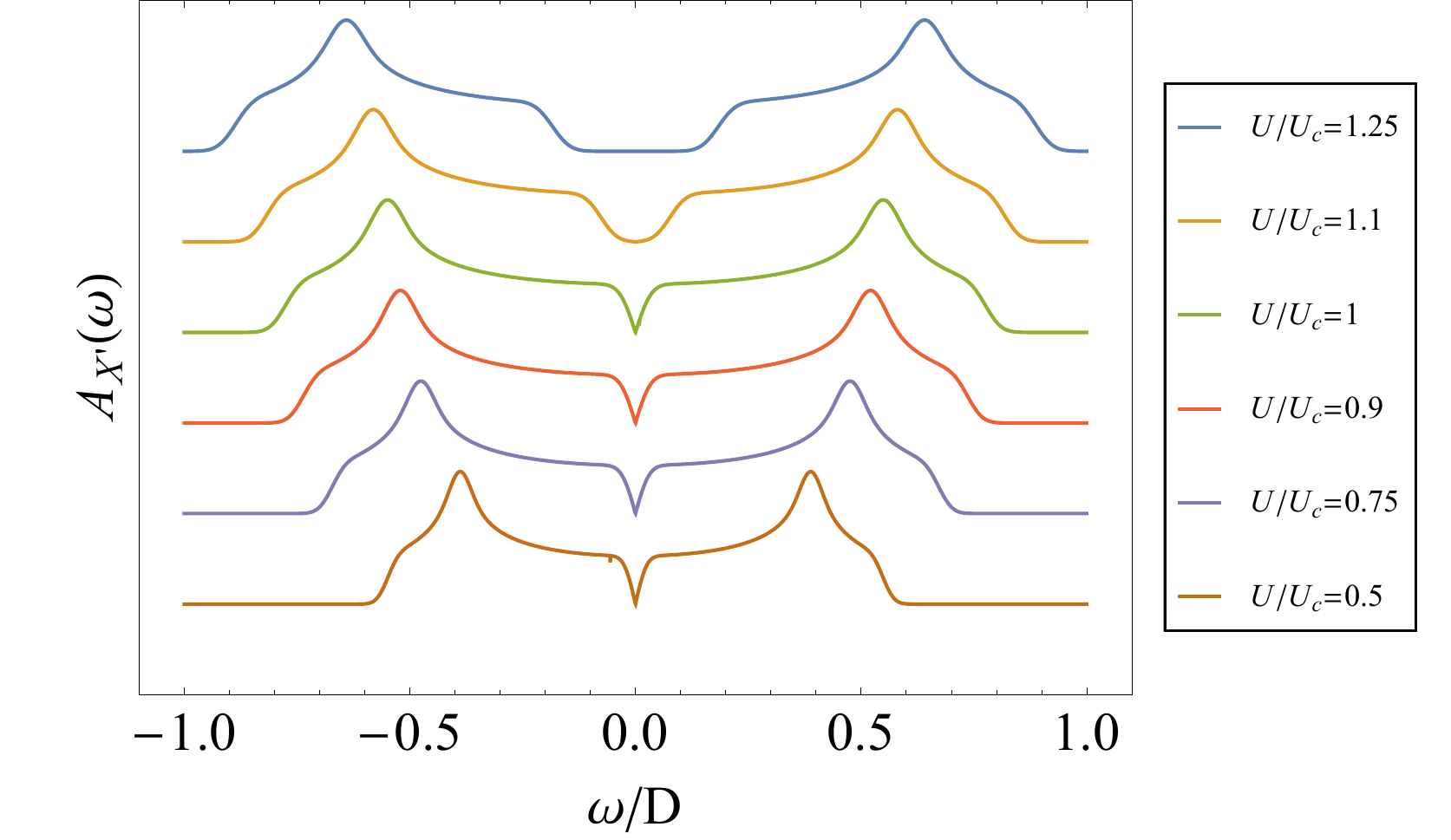}
  \caption{(Color online) The incoherent spectral function of the slave rotor field plotted versus $\omega/D$ for
  $U / U_c = 1.25, 1.1, 1, 0.9, 0.75, 0.5$ respectively. Here, $D$ is the electron bandwidth $D = 8 t$,
  and each curve is shifted
  by a different constant.
 When $U$ is larger than $U_c$, the rotor spectral weight has a gap around $\omega = 0$.
 In addition, it has two peaks
 near $\pm U \mathcal{O}(1)$, which correspond to two incoherent poles of the rotor Green's function.
 These poles persist into the bad metal regime across the Mott transition, as can be seen from the results
 for $U/U_c <1$.}
\label{fig:rotor-spectral-function}
\end{figure}

We therefore expect that $J_{\text{ex}}$ will be maximized around $U_c$. This is indeed seen in the calculated result
near the Mott transition, shown in Fig. (\ref{fig:JeffU}).

{\it The case of multiple orbitals %Application to the iron pnictides
}:
In iron pnictides, the atomic part includes the Hund's rule coupling. For simplicity, we consider the minimal two-orbital model\cite{Raghu2008b}
\begin{equation}\label{eq:two-orb-model}
  \begin{split}
     H_{at,2orb} &= \sum_i \big(U \sum_s n_{is \ua} n_{is\da} + (U' - \frac{J_H}{2})
    \sum_{\sigma,\sigma'}n_{i1\sigma} n_{i2\sigma'} \\ &  - 2 J_H \S_{i1}
    \cdot \S_{i2}\big),
  \end{split}
\end{equation}
where $U' = U - 2 J_H$ denotes the intraorbital Coulomb repulsion\cite{Castellani1978}.

In iron pnictides, $J_H$ is a fraction of $U$, thus a relevant energy scale in the atomic limit. Therefore, we solve the atomic energy levels with the $J_H$ term via exact diagonalization, and match the effect of Hund's rule coupling in the atomic within the above slave rotor framework.

At half-filling, the atomic ground state(s) are the spin-$1$ triplets states, with an charge excitation energy $E(N=3) + E(N=1) - 2 E(N=2)_{S=1} = U + J_H$. Near the half-filled insulating state, it is the superexchange interactions among the spin-$1$ local or quasi-local moments that are of interest.
In Ref. [\onlinecite{Meetei2013}], a renormalization of $U \rightarrow \tilde{U} = U + J_H$ is used to account for the correct Mott gap. However, the effect of Hund's coupling on spins is not accounted for, i.e. the formation of the spin-$1$ local moments. We further introduce a spinon Hund's coupling, which provides the correct spin-$1$ triplets when diagonalized in the atomic limit. 
%Due to the spin-charge separation nature of 
Within the slave rotor construction, the spinon Hund's coupling does not affect the rotor spectrum. Hence, we account for the correct spin-$1$ moments, as well as the correct charge excitations.
However, we must note that the contribution of Hund's coupling to the ground state energy is double-counted, and the Green's function is not as good. This is because the mapping only accounts for the first excited state in contrast to the $SU(2)$ symmetric case where the full spectrum is described by the rotor field.

Then the superexchange interaction $J_{\text{eff}}$ is computed in the same way via the two orbital slave rotor model with the renormalized $\tilde{U}$, either between the spinons, or the spin-$1$ operators. We always find that $J_{\text{eff}}$ peak at $U_c$, even though the value of $U_c$ is modified due to $J_H$. Numerical results for fixed $J_H = 0.2 U$ among spinons are plotted as functions of $U$ in Fig. (\ref{fig:JeffU}). %These results are compared with the second order perturbation results in Supplemental Material.
%\begin{figure}
% % \subfigure[]{\includegraphics[width=8.5cm]{./compare.pdf}\label{subfig:compare1}}
% % \subfigure[]{\includegraphics[width=8cm]{./compare2.pdf}\label{subfig:compare2}}
%  \includegraphics[width=7.5cm]{./Fig-5.pdf}
%  \caption{(Color online) $J_{\text{eff}}$ computed in limit ii) as a function of $U$ with $J_H = 0.2 U$. We find that $J_{\text{eff}}$ peaks at $U_c$ as the single orbital case.
%}\label{fig:Jeff-JH}
%\end{figure}
%
These interactions are more complex, because the appropriate spinon bilinears may not only involve the spin degrees of freedom, but also the orbital ones. The form of the latter will depend on the filling factors.
Here we will concentrate on the interactions that involve only the spins, in which case
the effective interaction will be the on-site Hund's couplings and intersite exchange interactions of the following form \cite{Si2008}:
$\sum_{ij,\tau\tau'}  J_{ij}^{\tau\tau'}  {\bf S}_i^{\tau} \cdot {\bf S}_j^{\tau'}$,
where $\tau,\tau'$ label the orbitals. The exchange interaction forms a matrix in the orbital basis.

{\it Discussions:~}
%{\it Generalization to Multi-Orbital Cases:~}
%The slave-rotor method can also be generalized to the multi-orbital cases \cite{Florens2004,Ko2011a}. From integrating out the incoherent spectra of the slave rotors, we will again get four-spinon interactions.
%The individual matrix element,$J_{ij}^{\tau\tau'}$, can be calculated within the framework we have presented here, and will depend on $U$ in a way similar to our results presented above for the single-orbital case.
We have used the slave rotor approach here as a means to capture the incoherent part of the electron spectral weight.
An attractive alternative approach is a slave-spin method, either in the Z$_2$ form \cite{Medici2005}
or in the $U(1)$ form \cite{Yu2012c}. The $U(1)$ formulation,
in particular, properly describes the Mott insulating phase and should therefore be able to capture the incoherent
part of the electron spectral weight in a similar fashion. This approach could be advantageous
to understanding the dependence of the exchange interactions on the Hund's
coupling \cite{Bascones2012b}.
Nonetheless, our procedure presented here already provides the conceptual basis for deriving the exchange interactions
in the bad-metal regime (in other words, not just on the insulating side).

 Although beyond the saddle point, a strongly coupled gauge theory is found \cite{Lee2005},  the high energy part of the spectrum does {\it not} involve a strongly coupled gauge theory. Provided $U/t$ is not too small (i.e., provided we stay in the bad-metal regime), the effect is not important. The calculation of the exchange interaction can be done with a low-energy cutoff in the integration over the slave-rotor spectral functions, and the result is insensitive to this cutoff.

Finally, we have focused on the 
%Heisenberg
bi-linear
 exchange interaction. Our procedure also contains processes for multi-spin exchange interactions. 
 These include both the bi-quadratic couplings
 and ring-exchange interactions.

{\it Implications for the Iron-based Superconductors:~}
As already discussed in the introduction, all iron-based superconductors
fall in the bad-metal regime. 
They are also inherently multi-orbital systems with a sizable ratio of the Hund's coupling to the Hubbard $U$, $J_H/U$.
Through our analysis of the multi-orbital Hubbard model with a ratio of $J_H/U$ comparable to that for the iron-based superconductors,
our work sheds light on the exchange interactions in these systems.

We note that the exchange interaction we have derived operates between the physical spin degrees of freedom. 
Through Eq.~(\ref{eq:d_by_fX}) and  Eq.~(\ref{eq:GX-MF-metallic}), we see that the single-electron spectral weight
contains a coherent part, with weight $Z$, and an incoherent part, with weight $1-Z$.
In the bad-metal regime, $Z$ is small. Therefore, the physical spin degree of freedom 
is predominantly generated by the incoherent single-electron excitations with the remainder contribution coming 
from the small coherent single-electron excitations near the Fermi energy \cite{Si2009a,Dai2009}.
This is to be compared with the fully localized regime, where the coherent electron weight vanishes ($Z=0$)
and the spin degrees of freedom would be just local moments.

Finally,
superconductivity can be driven by the short-range exchange interactions.
%Recent work
Work
by some of the co-authors
here \cite{Yu2013}
 have shown that the pairing amplitude increases with increasing $J_{\text{ex}} / D^*$, where $D^*$
 is the renormalized bandwidth.
 Since $D^*$ decreases as the Mott transition is approached, and $J_{\text{ex}}$ is maximized there, the superconducting pairing amplitude is expected to be the largest near the boundary of localization and delocalization.

To summarize, we have presented a microscopic approach to determine the exchange interactions
in the bad metal regime where the
% electrons fractionalize into both itinerant quasiparticles as well quasi-local moments due correlations.
single-electron excitation weight mostly lies in the incoherent part.
From concrete calculations, we have demonstrated that
 the exchange interaction is the largest near the boundary of localization and delocalization. Correspondingly, superconductivity driven by short-range spin-exchange interactions is expected to have the largest pairing amplitude in such a regime.

 \begin{acknowledgements}
 
%This work was completed prior to the passing of one of the authors (E.A.), 
%with whom the other three authors (W.D., R.Y. and Q.S.) are grateful for having had the opportunity to collaborate.
This work was completed prior to the passing of one of the authors
(E.A.). The other three authors (W.D., R.Y. and Q.S.) are grateful to
have had the opportunity to collaborate with him.
We would like to acknowledge the useful discussions with E. Bascones, S. A. Kivelson and other participants
of the KITP program on ``Magnetism, Bad Metals and Superconductivity: Iron Pnictides and Beyond". This work has been supported in part by the U.S. Department of Energy, Office of Science, Basic Energy
Sciences, under Award No.\ DE-SC0018197 and the Robert A. Welch Foundation Grant No.\ C-1411
(W.D. and Q.S.).
R.Y. was partially supported by the National Science Foundation of China
Grant No. 11674392, and the Fundamental Research Funds for the
Central Universities and the Research Funds of Renmin University
of China.
All of us acknowledge the support provided in part by the National Science Foundation under Grant
No. NSF 
PHY-1748958
%PHY11-25915 
at KITP, UCSB. We also acknowledge the hospitality of
the Aspen Center for Physics (NSF Grant No. PHY-1607611, Q.S.\ and E.A.).
\end{acknowledgements}
%%%%%%%%%%%%%%%%%%%%%%%%%%%%%%%%%
%\newpage
\appendix
\section*{Appendix: Application to two-orbital model: role of Hund's coupling}

To account for Hund's rule coupling in various models that describe the iron pnictides, we follow Ref. \cite{Meetei2013} to map the Hund's rule coupling to renormalization of the repulsion $U \rightarrow \tilde{U}$. 

The atomic (local) part of the minimal two orbital model reads
\begin{equation}
  \begin{split}
    H_{int} & = \sum_i \big(U \sum_s n_{is \ua} n_{is\da} + (U' -
      J_H/2) n_{i1} n_{i2} \\ 
      & - 2 J_H \S_{i1} \cdot \S_{i2}\big).
  \end{split}
\end{equation}
%where $\hat{N}_i = \sum_{\sigma,s} n_{is\sigma}$ is the total particle number operator at site $\x_i$, $n_{is} = \sum_\sigma n_{is\sigma}$, $T_{iz} = \sum_{s,s'} d^\dagger_{is} \tau^z_{s s'} d_{is'} = n_{i1} - n_{i2}$, $\S_i = \S_{i1} + \S_{i2}$, and for this two orbital model $N = 4$. Then 
We diagonalize $H_{int}$, and show several typical energies and eigenstates at different filling factors in Table \ref{tab:1}.
\begin{table}[ht]
 \centering
  \caption{Typical eigenstates and the corresponding eigenenergies of $H_{\text{int}}$ obtained by exact diagonalization. The eigenstates are labeled by the four occupation numbers $n_{i,\sigma}$ of orbital $i=1,2$ and spin $\sigma$.}\label{tab:1} 
 \renewcommand{\arraystretch}{1.6}% Spread rows out...
 \begin{tabular}{|c|c|c|c|c|c|c|c|c|c|}
\hline
   $n_{1\ua}$ & $n_{1\da}$ & $n_{2\ua}$ & $n_{2\da}$       & $E$ \\   \hline
1          & 0          & 0          & 0          &     0 \\ \hline
\dots      &            &            &               &     \\ \hline
1          & 0          & 1          & 0         &    $ U'- J_H $ \\ \hline
1          & 0          & 0          & 1          &    $U' + J_H$ \\ \hline
1          & 1          & 0          & 0          &     $U$ \\ \hline
\dots      &            &            &              &     \\ \hline
1          & 1          & 1          & 0          &   $2 (U'-J_H/2) + U$  \\ \hline
\dots      &            &            &              &     \\ \hline
1          & 1          & 1          & 1             &  $2(U+2U' - J_H)$   \\ \hline
  \end{tabular}
\end{table}
 
%In order to make $\ev{\hat{N}_i} = 2$ the ground state, we also need to include the chemical potential $\mu \hat{N}_i$.
Our focus is superexchange interaction of the $\ev{\hat{N}_i} = 2$ states. It is easy to verify that the energy spacing for the relavant excitations is $\Delta E = (E_i(N=1) - E_i(N=2)) + (E_i(N=3) - E_i(N=2)) =  U + J_H$ for the $S = 1$ triplet states.  For the $S=0$ singlet state, this energy spacing is $\Delta E = U - 3 J_H$. %Exchange between $S = 1$ and $S = 0$ sector also exist, with energy spacing $\Delta E = U - 1/2 J_H$.
Near the Mott insulating phase, the $S = 1$ triplets are the ground state. Even though the low lying $S=0$ state has lower excitation energy, it cannot participate in the superexchange virtual process. To further account for the spin splitting, we also include a spinon Hund's coupling
\begin{equation}
  H_{f,J_H} = - 2 J_H \sum_i \S_{f, i1} \cdot \S_{f, i2},
\end{equation}
where $\S_{f, i o} = \sum_{\alpha \beta} \frac{1}{2} f^\dagger_{io\alpha} \bm{\sigma}_{\alpha \beta} f^\dagger_{io\beta}$.

Therefore, we can account for both the charge excitation energy as well as the spin splitting of the two orbital Hubbard with Hund's coupling via
\begin{equation}
\tilde{H}_{at} = \sum_{i,\alpha} f^\dagger_{i\alpha} (\pt_\tau  + h_i) f_{i\alpha} + \sum_i  (\tilde{U}  \hat{L}_i^2 / 2 + h_i \hat{L}_i) + H_{f,J_H},
\end{equation}
with $\tilde{U} = U + J_H$. Since $H_{f,J_H}$ preserves 
the spinon number, $h_i = 0$ still holds at half filling.
%, which is just manifestation of spin-charge-separation nature of the slave rotor approach. 

%%%%%%%%%%%%%%%%%%%%%%%%%%%%%%%%%
%\bibliography{C:/bibtex/library}
%\bibliography{C:/Users/wxding/SkyDrive/PhyDir/Projects/library}
%\bibliography{C:/Users/dwx0_000/OneDrive/PhyDir/Projects/library}

\end{document}